\documentclass[prl,amsmath,amsfont,amssymb,twocolumn,showpacs]{revtex4-1}
\usepackage{graphicx}
\usepackage{color}
\usepackage[english]{babel}
\usepackage{amsmath}
\usepackage{bm}
\usepackage{graphicx}
\usepackage{setspace}
\usepackage{pgfplots}
\usepackage{comment}
\usepackage{subfigure}

\begin{document}

\def\beq{\begin{equation}}
\def\eeq{\end{equation}}
\def\brho{{\boldsymbol \rho}}
\def\bchi{{\boldsymbol \chi}}
\def\ba{{\bf a}}
\def\bb{{\bf b}}
\def\bc{{\bf c}}
\def\bw{{\bf w}}
\def\bd{{\bf d}}
\def\bk{{\bf k}}
\def\bp{{\bf p}}
\def\bq{{\bf q}}
\def\br{{\bf r}}
\def\bv{{\bf v}}
\def\bx{{\bf x}}
\def\bz{{\bf z}}
\def\bG{{\bf G}}
\def\bR{{\bf R}}
\def\bH{{\bf H}}
\def\bF{{\bf F}}
\def\bJ{{\bf J}}
\def\bP{{\bf P}}
\def\la{\langle}
\def\ra{\rangle}
\def\calH{\mathcal{H}}
\def\calM{\mathcal{M}}
\def\calP{\mathcal{P}}
\def\calL{\mathcal{L}}
\def\calV{\mathcal{V}}
\def\calE{\mathcal{E}}
\def\calZ{\mathcal{Z}}
\def\p{\hat {\psi}} 
\def\pd{\hat {\psi}^{\dag}}
\def\grad{\mbox{\boldmath $\nabla$}}
\def\Tr{{\rm Tr}}
\def\e{\epsilon}
\def\ve{\varepsilon}
\def\pa{\partial}
\def\nn{\nonumber}
\def\t{\tau}
\def\kbar{\bar {k}}
\def\rbar{\bar {r}}
\def\nbar{\bar {n}}

\title{Liquid crystal phases of two-dimensional dipolar gases and Berezinskii-Kosterlitz-Thouless melting}
\author{Zhigang Wu, Jens K. Block and Georg M. Bruun}
\affiliation{Department of Physics and Astronomy, Aarhus University, DK-8000 Aarhus C, Denmark}
\date{\today}
\begin{abstract}
Liquid crystals are phases of matter intermediate between crystals and liquids. Whereas classical liquid crystals have been known for a long time and are used in electro-optical displays, much less is known about their quantum counterparts. There is growing evidence that quantum liquid crystals play a central role in many electron systems including high temperature superconductors, but a quantitative understanding is lacking due to disorder and other complications. Here, we analyse the quantum phase diagram of a two-dimensional dipolar gas, which exhibits stripe, nematic and supersolid phases. We calculate the stiffness constants determining the stability of the nematic and stripe phases, and the melting of the stripes set by the proliferation of topological defects is analysed microscopically. Our results for the critical temperatures of these phases demonstrate that a controlled study of the interplay between quantum liquid and superfluid phases is within experimental reach for the first time, using dipolar gases.  
\end{abstract}
\maketitle

The  investigation of cold atomic gases has  enabled one to study many-body physics with unrivalled experimental control and in regimes never realised before. Recent  progress in trapping and cooling of  gases consisting of dipolar atoms/molecules opens up a promising new research direction. The dipole-dipole interaction is long range and anisotropic, which is predicted to give rise to a number of exotic forms of matter~\cite{Baranov2008,Lahaye2009}. Degenerate Fermi gases consisting of atoms with a large magnetic dipole moment have already been created~\cite{Lu2012,Aikawa2014}, and progress towards producing degenerate gases of fermionic molecules with an electric dipole moment is being reported~\cite{Heo2012,Park2015}. 

Here, we analyse the quantum phases of a  two-dimensional (2D) dipolar Fermi gas at non-zero temperatures $T$. This includes a stripe phase, whose 
low energy degrees of freedom are described by an anisotropic XY model. We determine the stiffness constants of this effective model microscopically. 
The finite temperature melting of the stripe phase is driven by the proliferation of topological defects called dislocations, and the corresponding
 Berezinskii-Kosterlitz-Thouless (BKT) critical temperature is determined by the well-known renormalisation group equations. For large tilting angles of the dipoles, the system can have additional superfluid pairing which coexists with the stripe order. We calculate the critical temperature of the superfluid transition. When the dipoles are perpendicular to the 2D plane, the critical temperature of the stripe phase is shown to vanish, and the system exhibits a nematic phase characterised by long range orientational order but no translational order. 
Our results demonstrate that  with dipolar gases, it  is within experimental reach to study 
quantum liquid and superfluid phases characterised by varying degrees of spontaneous translational, rotational, and gauge symmetry breaking. The interplay 
between such  phases is believed to play an important role in many electronic materials 
discovered in recent decades~\cite {Fradkin2010,Kivelson1998,Chuang2010,Emery1999,Kohsaka2007}. Moreover, our results show that one can confirm the microscopic
 mechanism behind the BKT transition, namely the proliferation of topological defects, simply by observing the proliferation of disclocation defects in the stripe pattern. 
Such an experimental verification of BKT physics has been achieved only recently using atomic gases~\cite{Hadzibabic2006}, whereas other  experiments  reported only 
indirect evidence of BKT 
physics in the bulk properties~\cite{Bishop1978,Fiory1983,Reyren2007,Ye2010,Matthey2007,Rout2010,Resnick1981,Safonov1998}.

\section{Results}
We consider fermionic dipoles of mass $m$ and average areal density $n_0$, which are restricted to move in the  $xy$ plane by a tight harmonic trapping 
potential $m\omega_z^2z^2/2$ along the $z$-direction. In the limit $\omega_z \gg \e_F$, where $\e_F=k_F^2/2m=2\pi n_0/m$ is the Fermi energy of a 2D 
non-interacting gas with  areal density $n_0$, the system is effectively 2D with the dipoles  frozen in the harmonic oscillator ground state in the $z$ direction. An external field aligns the dipoles so that their  dipole moment $\bf d$ is perpendicular to the $y$-axis and forms an angle $\Theta$ with the $z$-axis. 
The dipole-dipole interaction is $V_{\rm d}(\br,z)  = D^2(1-3 \cos^2\theta_{rd})/(r^2+z^2)^{3/2}$, where  $\theta_{rd}$ is the angle between the relative displacement vector of the two dipoles $(\br,z)$ with $\br=(x,y)$ 
and the dipole moment $\bf d$, and $D^2 = d^2/4\pi\ve_0$ for electric dipoles and $D^2 = d^2\mu_0/4\pi$ for magnetic ones. 

The strength of the interaction is determined by the dimensionless parameter $g = 4mD^2k_{F}/3\pi$, and the degree of anisotropy 
is controlled by the tilting angle $\Theta$. The system is rotationally symmetric for $\Theta=0$ and becomes more anisotropic with increasing $\Theta$. Above a critical  interaction strength $g_c(\Theta)$, it is predicted to form density stripes at $T=0$, where the density exhibits periodic modulations of the form  
\begin{equation}
n({\mathbf r})=n_0+n_1\cos ({\mathbf q}_c\cdot{\mathbf r}-u). 
\label{Density}
\end{equation}
Here, $\bq_c\parallel{\mathbf e}_y$ is the wave vector of the stripes, and $n_1$ and $u$ are their amplitude and phase respectively. The density modulation 
is formed along the $y$-direction so as to minimise the interaction energy. The system thus exhibits liquid-like correlations 
along the $x$-direction and crystalline correlations along the $y$-direction. 
This phase has been predicted  by Hartree-Fock theory~\cite{Yamaguchi2010,Babadi2011,Sieberer2011,Block2012,Block2014}, density-functional theory~\cite{vanZyl}, and 
by a variant of the so-called STLS method~\cite{Parish2012}. Remarkably, Hartree-Fock and density-functional theory 
predict essentially the same critical coupling strength $g_c(0)\simeq0.6$ for stripe formation at $\Theta=0$, whereas the STLS method obtains a somewhat higher value. 
For $\Theta\gtrsim0.23\pi$, the system is predicted to become a $p$-wave superfluid~\cite{Bruun2008}, which for strong enough coupling can coexist with the stripe order 
forming a supersolid~\cite{ZhigangWu2015}. The dipoles are also predicted to form a Wigner crystal for  $g\simeq27$ for $\Theta=0$~\cite{Astrakharchik2007,Buchler2007}. This very strong coupling regime is outside the scope of the present paper. 

\subsection{Stripe phase at finite $T$ and effective XY model}
Since the stripe phase  breaks translational invariance along the $y$-direction, it is a quantum analog of a classical smectic liquid crystal~\cite{ChaikinBook,NelsonBook}. Indeed, the system  has a manifold of equivalent ground states distinguished only by a constant  factor $u$, which specifies the position of the stripes along the $y$-direction. 
Consequently, there are low energy collective excitations associated with a spatially dependent phase $u(x,y)$. Moreover, since a change from $u$ to $u+2\pi$ returns 
the system to the same ground state,  it follows that the low energy degrees of freedom of the stripe phase are described by a 2D anisotropic XY model. Specifically,  the simplest form of the elastic free energy congruent with the symmetry of the system is given by 
\begin{equation}
F_{\rm el}
= \frac12\int\! d^2r[B_\perp(\partial_xu)^2+B_\parallel(\partial_y u)^2]=\frac B 2  \int\! d^2r(\nabla u)^2
\label{Fel}
\end{equation}
for $\Theta \neq 0$. Here, $B_\perp$ and $B_\parallel$ are the perpendicular and parallel elastic coefficients describing respectively 
the energy cost of small rotations and compressions/expansions the stripes. In the second equality, we have used the rescaling $x\rightarrow \sqrt{B_\parallel/B_\perp}x$ to obtain an isotropic XY model with the effective elastic constant  $B = \sqrt{B_\perp B_\parallel}$. 

\subsection{Berezinskii-Kosterlitz-Thouless melting}
\begin{center}
\begin{figure}
\includegraphics[width=0.75\columnwidth]{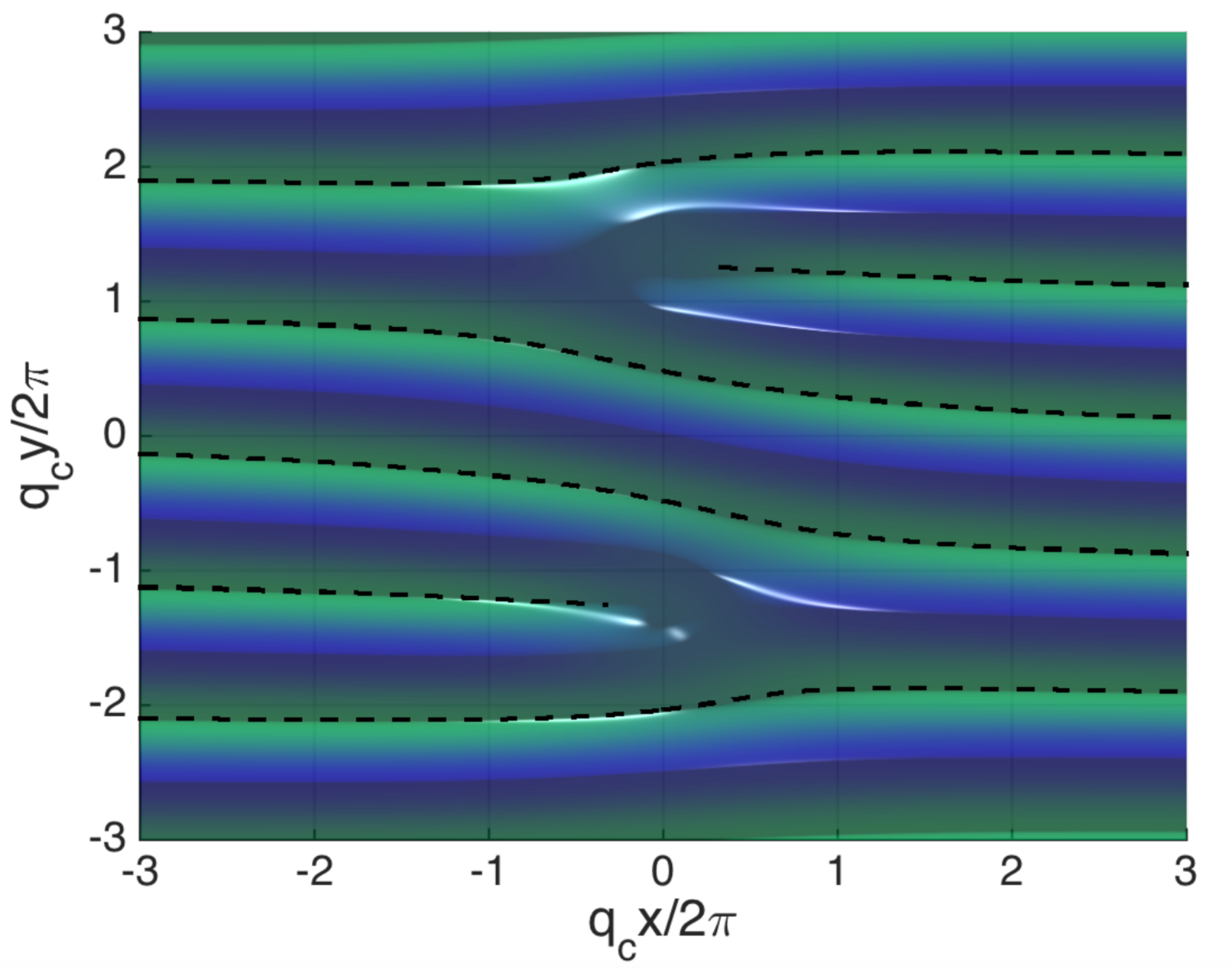}
\includegraphics[width=0.75\columnwidth]{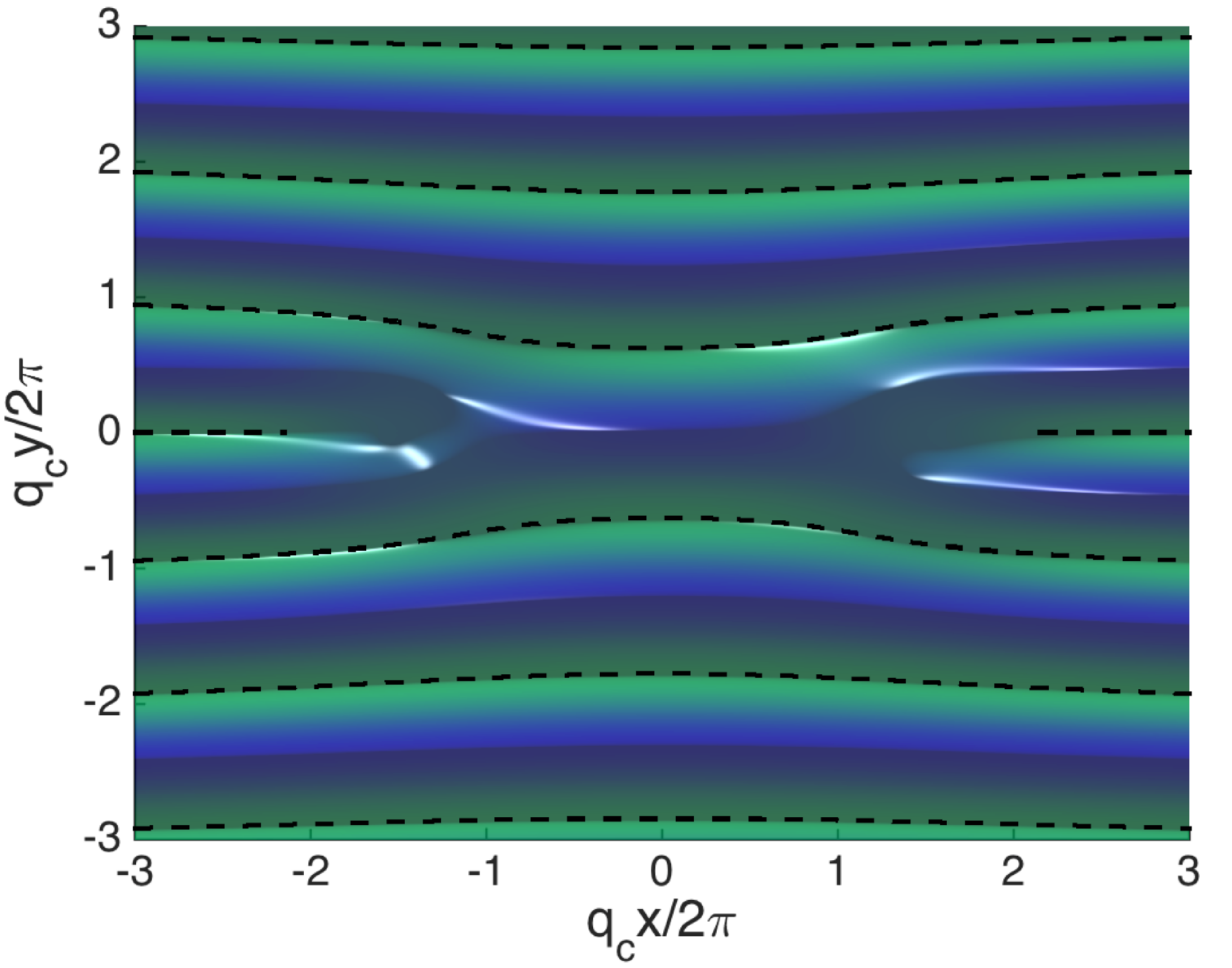}
\caption{\textbf{Dislocation pairs in the stripe phase.} The dislocations are centered at  $(x_1,y_1)$ and $(x_2,y_2)$ so that the phase field is $u({\bf r})=\arctan[(y-y_1)/(x-x_1)]-\arctan[(y-y_2)/(x-x_2)]$. Top: A $Q=1$ dislocation centered at $(0,-2.7\pi/q_c)$ and a $Q=-1$ dislocation centered at $(0,2.7\pi/q_c)$. Bottom: A $Q=1$ dislocation centered at $(-3\pi/q_c,0)$ and a $Q=-1$ dislocation centered at $(3\pi/q_c,0)$. The dashed lines 
indicate the position of the density maxima. } 
\label{dpair}
\end{figure}
\end{center}

As the stripe phase is described by the XY model, it exhibits algebraic long-range order at sufficiently low temperatures and it melts via the Berezinskii-Kosterlitz-Thouless mechanism due to the proliferation of topological defects~\cite{Berezinskii1972,Kosterlitz1973,Kosterlitz1974,Jose1977}. In the case of the stripe phase, the topological defects are dislocations. The phase field for a  single dislocation of charge $Q=\pm1, \pm2\ldots$ satisfies 
$\oint \nabla u(\br)\cdot d {\bf l} =2\pi Q$, where the path of the integration encloses the core of the dislocation. The presence of such a dislocation corresponds to inserting $Q$ extra stripes to the left (right) of the dislocation for $Q>0$ ($Q<0$). The energy of a single dislocation consists of a core part $E_c$,  and a part that scales logarithmically with the size of the system. Pairs of bound dislocations with opposite charges $Q=\pm1$ ($|Q|>1$ are energetically suppressed), 
however, have a finite energy even for an infinite system size and can be thermally excited in the stripe phase. This is due to the fact that the phase fields of the oppositely-charged dislocations cancel at large distances,
 which results in merely a local disturbance of the density. In Fig.\ \ref{dpair}, we illustrate dislocation pairs with opposite charges $Q=\pm1$ centered at $(x_1,y_1)$ and $(x_2,y_2)$ respectively. The stripe amplitude is suppressed in the core regions of the defects due to the large 
energy cost associated with $\nabla u\propto 1/r$, where $r$ is the distance to the core. From the rescaling $x\rightarrow \sqrt{B_\parallel/B_\perp}x$ it follows that the energy of a vertically displaced dislocation pair distance $\delta$ apart is the same as that of a pair displaced horizontally by the distance $\sqrt{B_\perp/B_\parallel}\delta$. Since $B_\parallel>B_\perp$ as we will demonstrate below, this shows that the dislocation pairs along the $x$-direction are more tightly bound than those along the $y$-direction. 

The spontaneous thermal excitation of bound dislocation pairs decreases the elastic coefficients at a macroscopic scale. The softening of the effective stiffness constant $B$ can be calculated from the well-known renormalisation group equations as described in the methods section. At a critical temperature $T_c^\text{st}$, the  renormalised elastic coefficient $B_R$ drops to zero by a sudden jump of magnitude $2T_c^\text{st}/\pi$. This disappearance of elastic rigidity signals the melting of the density stripes.

\subsection{Calculation of bare stiffness constants}
\begin{center}
\begin{figure}[ht]
\includegraphics[width=0.49\linewidth]{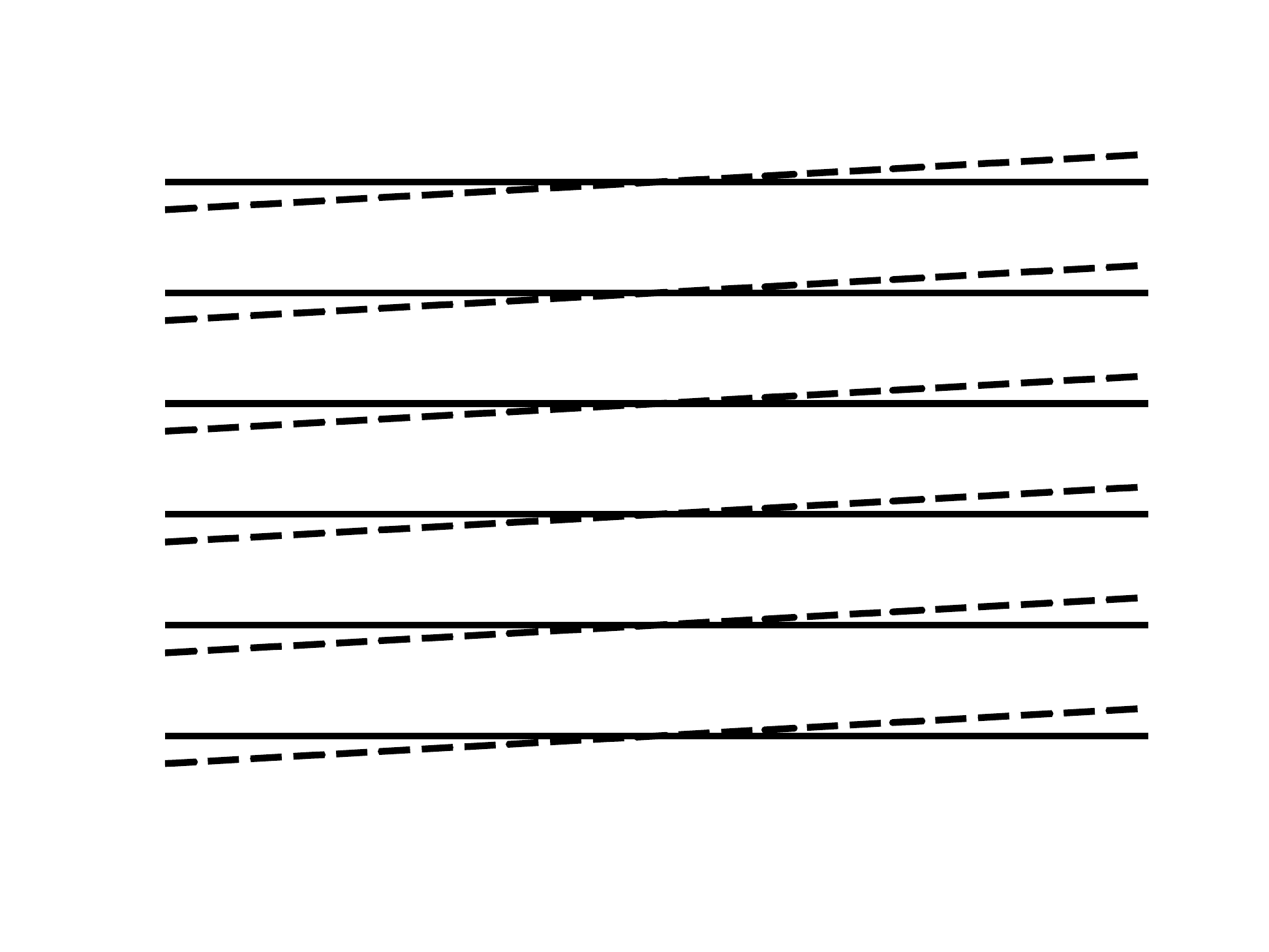}
\includegraphics[width=0.49\linewidth]{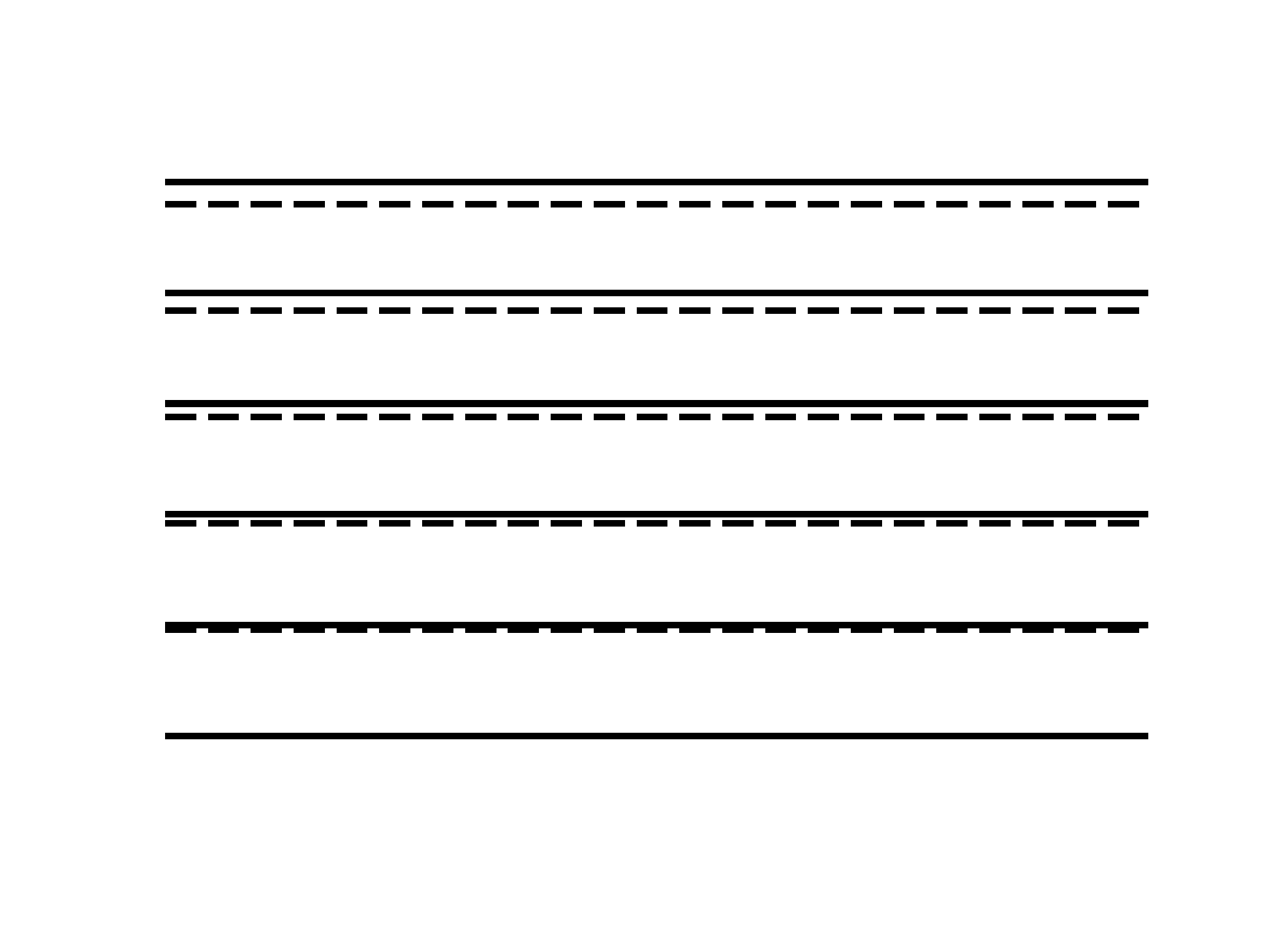}
\caption{\textbf{Elementary distortions of the stripes.} Rotation (left) and compression (right) of the stripes away from their equilibrium positions  indicated by the solid lines.  }
\label{ew}
\end{figure}
\end{center}

We now turn to a microscopic calculation of the ``bare" stiffness constants $B_\parallel$ and $B_\perp$ unrenormalised by dislocation pairs. The relevant thermodynamic quantity is the free energy of the system $F(\bq)$, which depends  on the stripe wave vector $\bq$.  Any non-uniform phase fluctuation increases the free energy by an amount given by (\ref{Fel}) for long wave lengths. To extract the elastic coefficients $B_\perp$ and $B_\parallel$, we consider two specific  distortions: an infinitesimal rotation and an infinitesimal compression/expansion of the stripes away from the equilibrium configuration, as illustrated in Fig.~\ref{ew}. 
These distortions are described by the phase field $u({\mathbf r})=-\delta{\mathbf q}\cdot{\mathbf r}$, where $\delta{\mathbf q}=\delta q_\parallel\cdot\mathbf{e}_y$ for the compression and $\delta{\mathbf q}=\delta q_\perp\cdot\mathbf{e}_x$ for the rotation. They are thus equivalent to a variation of the stripe vector  ${\mathbf q}={\mathbf q}_c+\delta{\mathbf q}$. Inserting the phase fluctuations into Eq.~(\ref{Fel}), we obtain the increment of the free energy 
\begin{align}
F_{\rm el}  = F(\bq_c+\delta \bq) - F(\bq_c)  
 = \frac A2(B_\perp\delta q_\perp^2+B_\parallel\delta q_\parallel^2),
 \label{Felexp}
 \end{align} 
where $A$ is the area of the system, and we have used the equilibrium condition $\nabla F(\bq)|_{\bq_c}=0$.  We thus find
\begin{equation}
B_\perp=\frac 1 A\left.\frac{\partial^2 F(\bq)}{\partial q_\perp^2}\right |_{\bq_c},  \quad
B_\parallel=\frac 1 A\left.\frac{\partial^2 F(\bq)}{\partial q_\parallel^2}\right |_{\bq_c}.
\label{Bcoef}
\end{equation}
The interaction energy per particle due to stripe formation scales as $D^2n_0^{-3/2}(n_1/n_0)^2$. Assuming that the interaction energy is dominant, we find that the elastic coefficient $B$ scales as $\sim (n_1/n_0)^2g\epsilon_F$ for a fixed $\Theta$. The magnitude of $B$ can be further reduced by a geometrical factor depending on $\Theta$, since the system becomes 
rotationally symmetric for $\Theta=0$, as we shall discuss below.

In order to microscopically calculate  the bare stiffness constants, we employ Hartree-Fock mean-field theory for the free energy, writing $F(\bq)\simeq F_\text{MF}(\bq) = \Omega_\text{MF} + \mu N$, where $\Omega_\text{MF}$ is the mean-field thermodynamic potential given by 
\begin{align}
\Omega_\text{MF} = -  E_{\rm int}- T\sum_{j,\bar\bk} \ln\left [ 1+ e^{-\beta \left(\varepsilon_{j\bar\bk}-\mu\right )}\right ],
\label{OmegaMF}
\end{align}
$\mu$ is the chemical potential and $N$ is the total number of particles. The quasiparticle energies are $\varepsilon_{j\bar\bk}$, where 
$j=1,2,\cdots$ is the band index and $\bar\bk$ is restricted to the first Brillouin zone of the 1D periodic potential set up by the stripes.
We subtract  the interaction energy $ E_{\rm int}$ to avoid double counting. The details of this calculation are given in the methods section. 

\begin{center}
\begin{figure}[ht]
\includegraphics[width=0.99\linewidth]{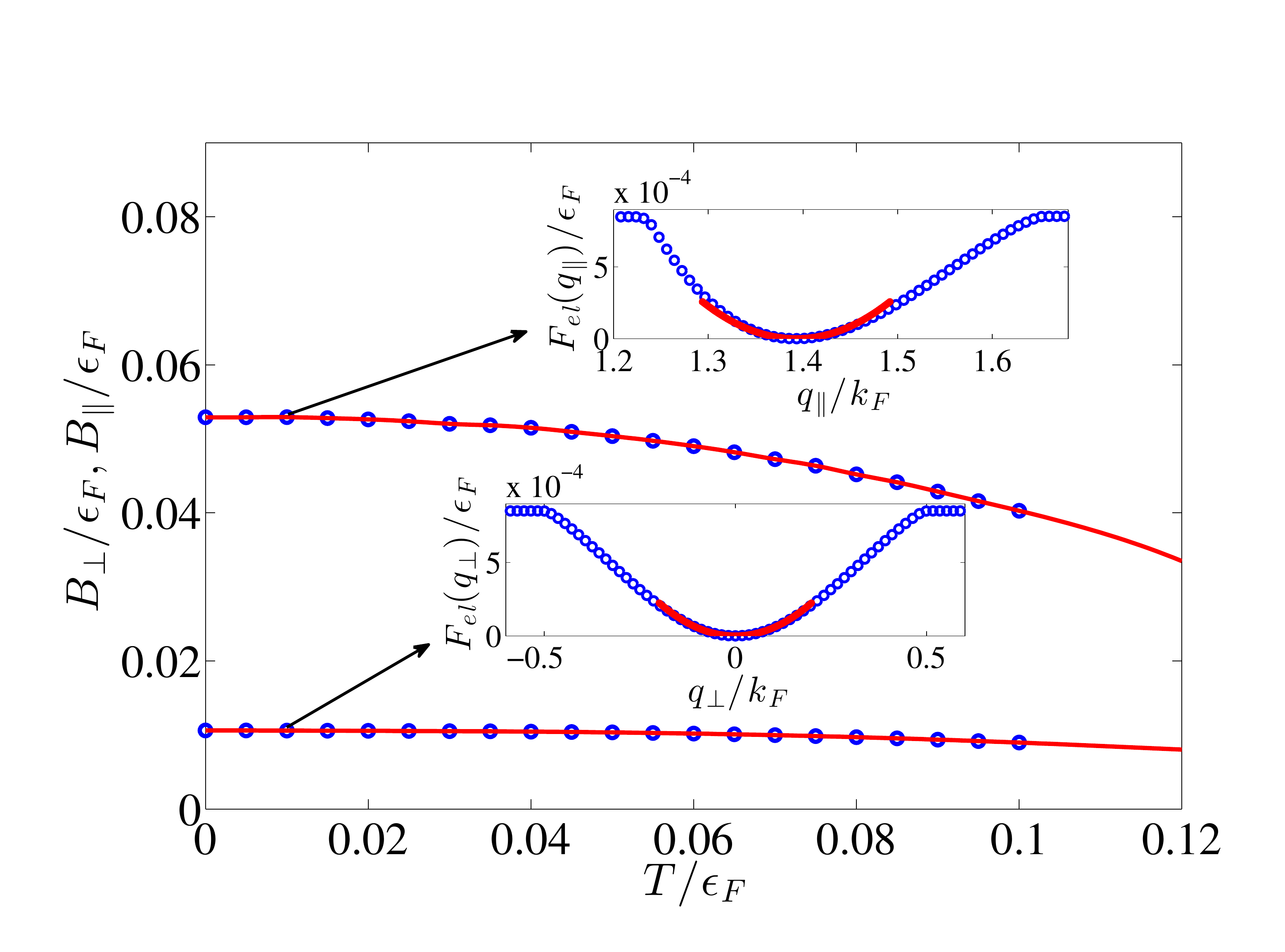}
\caption{\textbf{Bare stiffness constants as a function of temperature.} Upper curve gives $B_\parallel$  and lower curve $B_\perp$  for $g=1$ and $\Theta = 0.28\pi$. The blue circles in the insets are elastic free energy plotted as a function of $\bq$ at $T = 0.01\epsilon_F$, and the red solid curves are parabolic fits to several data points in the vicinity of $\bq = \bq_c$.}
\label{BT}
\end{figure}
\end{center}

In Fig.~\ref{BT}, we plot the bare elastic coefficients obtained from this approach as a function of  temperature for $g = 1$ and $\Theta = 0.28\pi$, for which the system has a large stripe amplitude $n_1/n_0 \simeq 0.6$ at low temperatures. In order to minimize finite size effects, we determine the elastic coefficients  by fitting a parabolic curve to the free energy in the vicinity of $\bq = \bq_c$ in accordance with (\ref{Felexp}), instead of performing a numerical differentiation following (\ref{Bcoef}). This is illustrated in the insets of Fig.~\ref{BT}. This procedure allows us to obtain numerically accurate values for the elastic coefficients. 
From  Fig.~\ref{BT}, we see that both elastic coefficients decrease with increased temperature. This is expected 
since thermal excitations of quasi-particles reduce the stripe amplitude and thus their rigidity. We also find that  $B_\perp\ll B_\parallel$, which suggests that 
 compressing/expanding the stripes costs more energy than a rotation. This difference in magnitude becomes even more profound for small $\Theta$ when $B_\perp$ is strongly suppressed by the weak anisotropy of the system. Finally we note that for $g = 1$ and $\Theta = 0.28\pi$, the system is in fact predicted to have additional superfluid pairing at $T = 0$~\cite{ZhigangWu2015}.  However, as demonstrated in Ref.~\cite{ZhigangWu2015}, the superfluid order has 
negligible effects on the stripe formation, and it can thus be safely neglected when analysing the elastic properties of the stripes. 

\begin{center}
\begin{figure}[ht]
\includegraphics[width=0.99\linewidth]{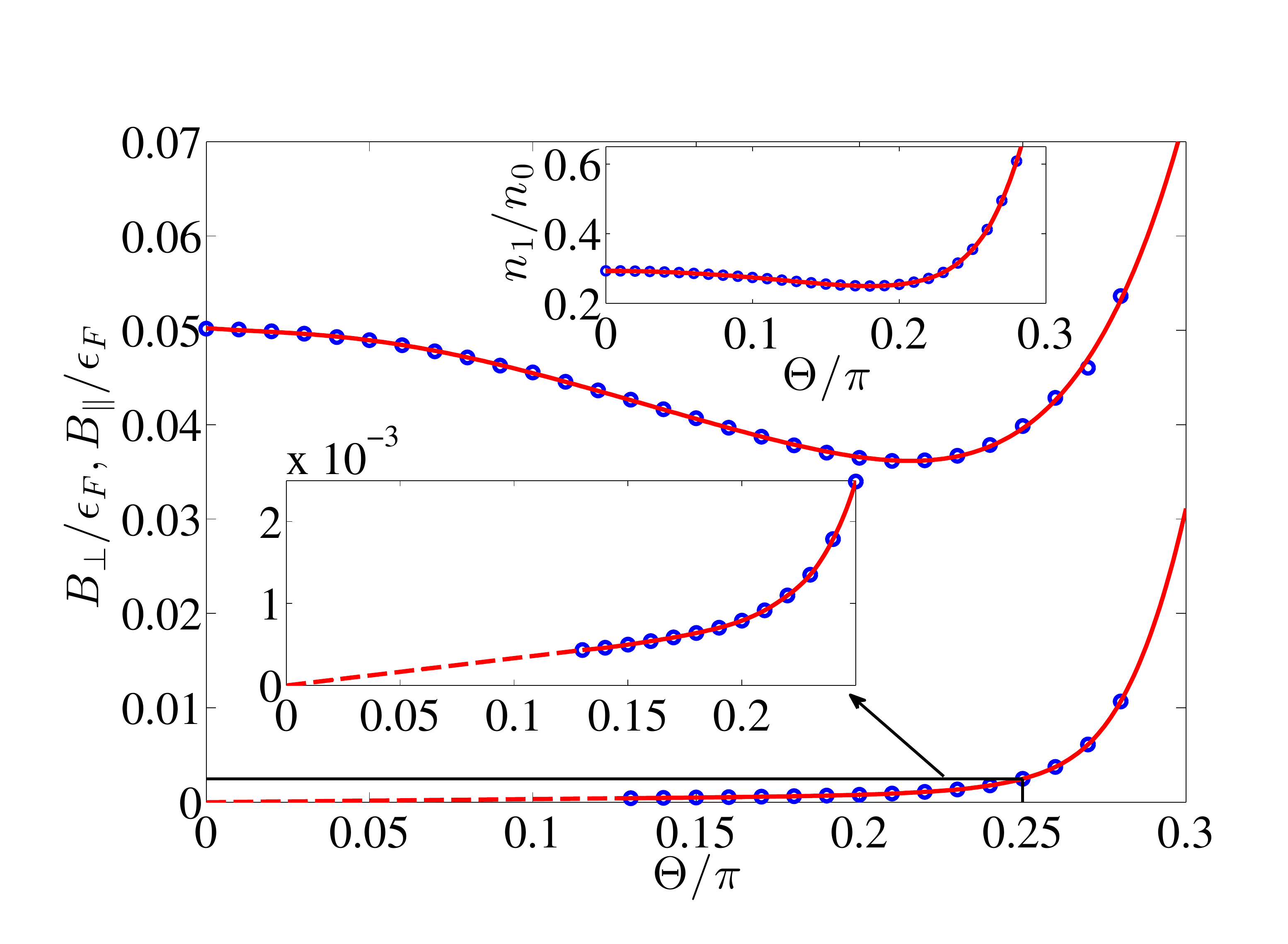}
\caption{\textbf{Bare stiffness constants as a function of tilting angle.} Upper curve gives $B_\parallel$  and lower curve $B_\perp$   for $g=1$ and $T = 0.01\epsilon_F$. For $B_\perp$, the red dashed line is an extrapolation for $\Theta<0.13\pi$, where the coefficient is too small to be accurately determined with our numerical method. The upper inset is a plot of the relative stripe amplitude as a function of $\Theta$. The lower inset is an expanded view of the $B_\perp$ for small values of $\Theta$. }
\label{B_Theta}
\end{figure}
\end{center}

In  Fig.~\ref{B_Theta}, we plot the bare elastic constants as a function of the tilting angle $\Theta$ for $g = 1$ and $T=0.01\epsilon_F$. 
The elastic constant $B_\parallel$ depends non-monotonically on  $\Theta$, first decreasing and then increasing exhibiting a minimum at 
$\Theta\simeq0.24\pi$. This is consistent with the mean-field phase diagram, which shows that the stripe formation is somewhat suppressed for 
intermediate values of $\Theta$~\cite{Block2014,ZhigangWu2015}. To illustrate this, we plot as an inset the stripe amplitude $n_1$ as a function of $\Theta$; we see that it exhibits the same non-monotonic behaviour as $B_\parallel$. In comparison to this behaviour, Fig.~\ref{B_Theta} shows that $B_\perp$ increases monotonically in $\Theta$. In particular, we have $B_\perp\rightarrow0$ for $\Theta\rightarrow0$ as shown in detail in the inset. This 
reflects that the system is rotationally symmetric for $\Theta=0$ such that a rotation of the stripes costs no energy.

\subsection{Renormalised stiffness constants and stripe melting}
\begin{center}
\begin{figure}[ht]
\includegraphics[width=0.99\linewidth]{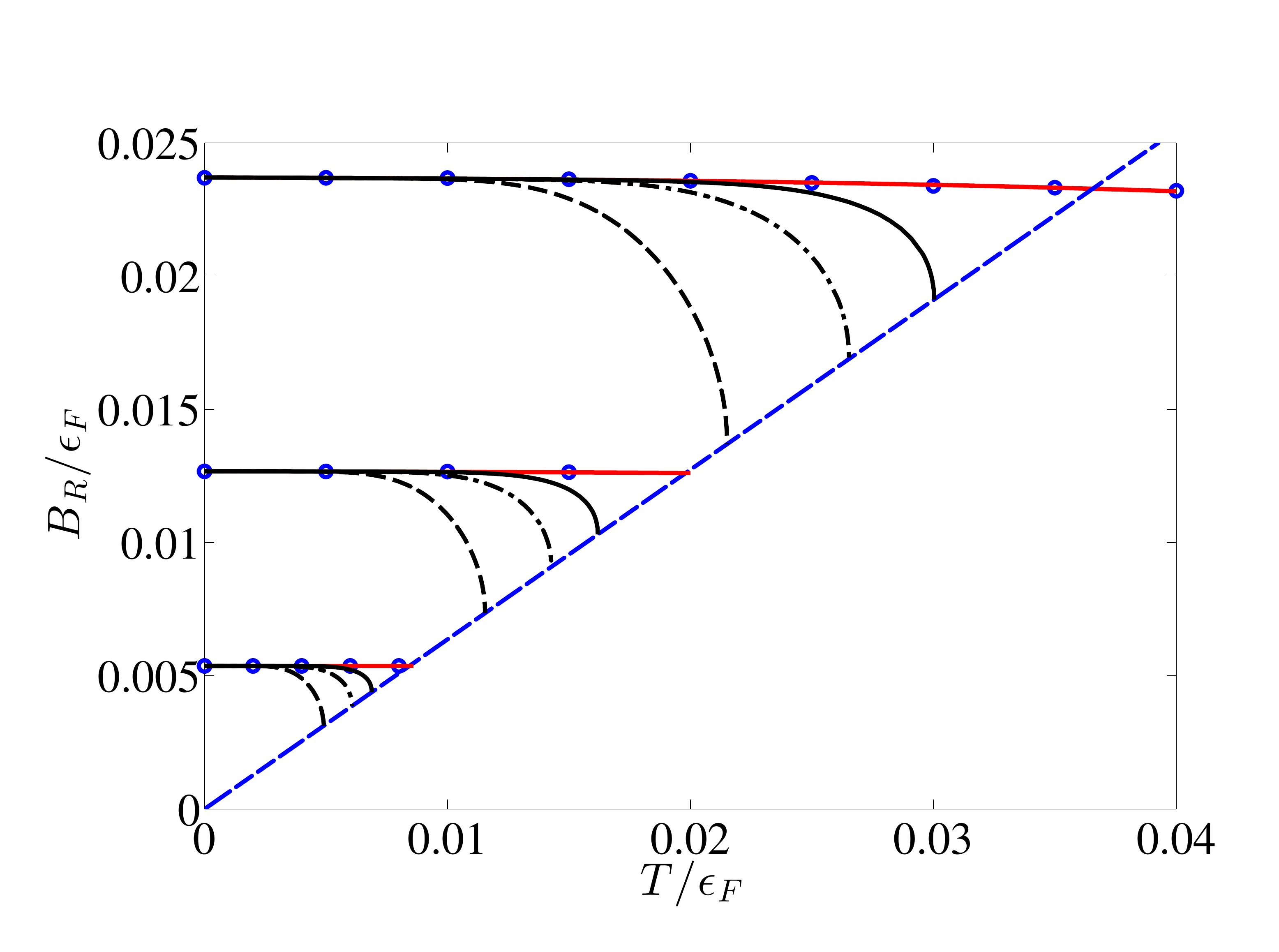}
\caption{\textbf{Renormalised stiffness constants as a function of temperature.} The three groups of curves for $B_R(T)$ for $g = 1$ correspond to, in the order from bottom to up,  $\Theta = 0.2\pi$, $\Theta = 0.26\pi$ and $\Theta = 0.28\pi$ respectively. In each of the group, the four curves, in the order from bottom to up, correspond to $\kappa = 2,3,4$ and $\infty$ (mean-field result) respectively. The slope of the dashed diagonal line is $2/\pi$. }
\label{BR}
\end{figure}
\end{center}

The bare elastic constants obtained from the mean-field theory can now be used as initial values in the RG equations to determined the renormalised elastic constants. We also need  the dislocation core energy, which  must scale as $\sim B$. Therefore, we  write $E_c = \kappa B$, where $\kappa$ is a constant of order unity. In Fig.~\ref{BR}, we plot the renormalised elastic coefficient $B_R$ as a function of temperature, obtained by solving (\ref{RGeqn}) with the initial mean-field values of $B=\sqrt{B_\perp B_\parallel}$ and $E_c = \kappa B$ for various coupling strengths $g$ and tilting angles $\Theta$. To examine the dependence on the core energy, we have chosen different values of $\kappa$. We see that the thermal excitation of dislocation pairs soften the elastic coefficients as expected. This softening is negligible for low $T$ where the core energy prohibits the excitation of dislocations. The softening increases with decreasing core energy and increasing $T$. At the critical temperature $T_c^\text{st} $ determined by the solution to $2T/\pi= B_R(T)$, the elastic coefficient drops to zero discontinuously and the stripes melt. 

\begin{center}
\begin{figure}[ht]
\includegraphics[width=0.99\linewidth]{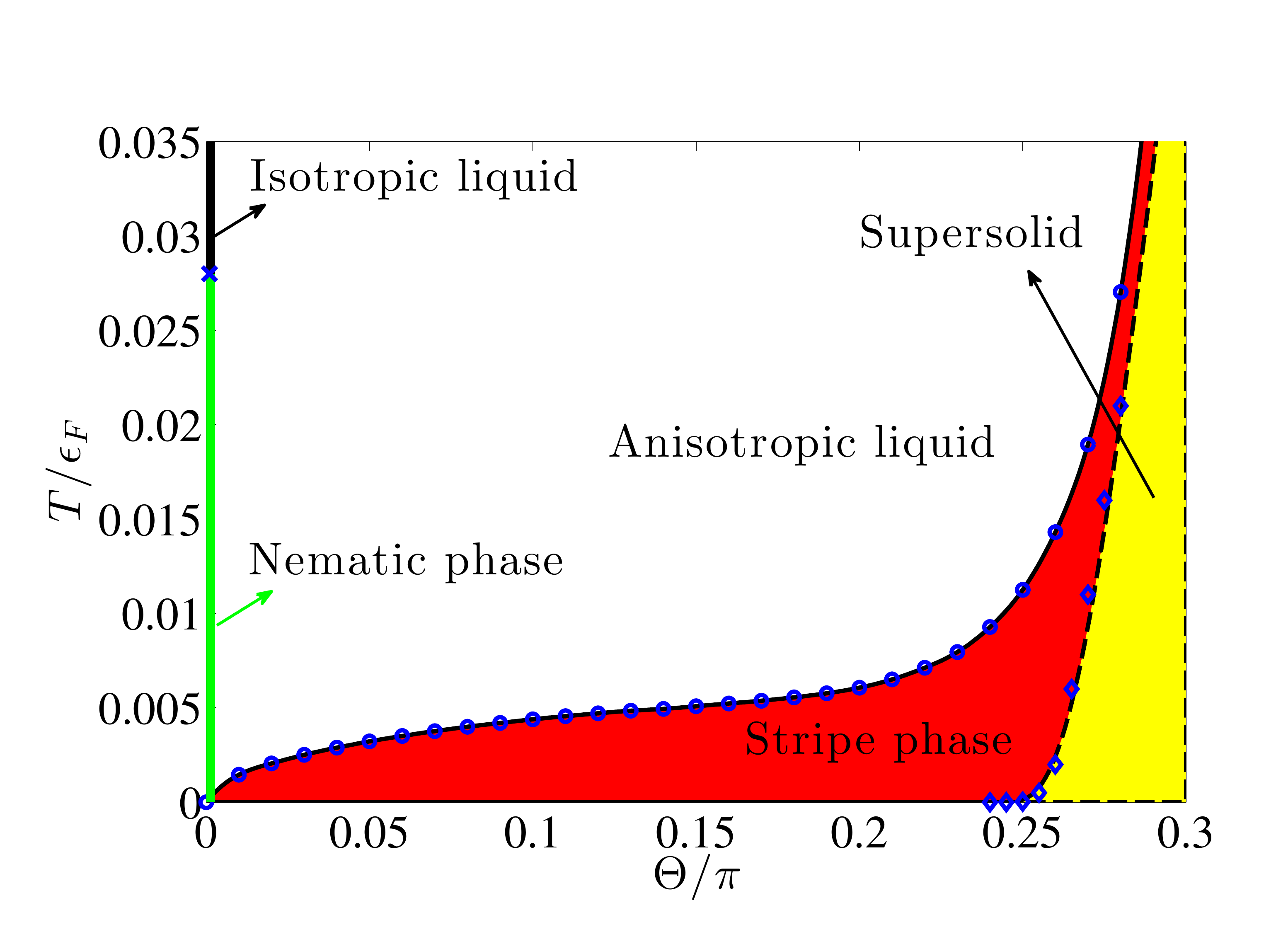}
\caption{\textbf{Phase diagram for $g = 1$.} The system is in the stripe phase below the Kosterlitz-Thouless melting temperature which is calculated taking $\kappa = 3$ for the core energy. For $\Theta=0$, the striped phase melts at $T=0_+$ into a nematic phase with long range orientational order but no translational order. The nematic phase melts into an isotropic liquid at a temperature $T_c^\text{n}\simeq B_\parallel$ indicated by the blue cross. For $\Theta\gtrsim\arcsin(2/3)$ the system is in a supersolid phase at $T=0$ 
with both stripe and superfluid order. The transition temperatures of this  phase calculated for various tilting angles are indicated by the diamonds. The dashed curve is a fit to the data by (\ref{Ts}).} 
\label{Tctheta}
\end{figure}
\end{center}

The resulting melting temperature $T_c^\text{st}$ is plotted in Fig.\ \ref{Tctheta}  as a function of  $\Theta$ for  $g=1$ and $\kappa = 3$. It increases rapidly with $\Theta$, indicating that the degree of anisotropy of the system increases such that the stripes become more rigid.  An extrapolation of our calculation for $g = 1$ and $\kappa = 3$ shows that  $T_c^\text{st} \sim 0.06 \e_F$ for $\Theta\simeq 0.3\pi$. The critical temperature 
also increases with the coupling strength, scaling as $T_c^\text{st}\sim B\sim (n_1/n_0)^2g\epsilon_F^0$. We note that in addition to the explicit linear dependence on $g$, the $T_c^\text{st}$ can further increase with the coupling strength through the dependence on $n_1$.  Our results show that 
in order to observe the stripe phase and the associated BKT physics with dipoles, it is preferable to choose a large tilting angle in addition to having a large dipole moment. However, the tilting angle cannot exceed $\Theta\simeq 0.3\pi$ above which the system exhibits a density collapse for large coupling strengths~\cite{Bruun2008,ZhigangWu2015}.

\subsection{Melting of supersolid phase}
The system exhibits $p$-wave pairing for $\Theta > \arcsin(2/3)$~\cite{Bruun2008}, which can coexist with stripe order for $g> g_c(\Theta)$ 
at $T=0$~\cite{ZhigangWu2015}. We now determine the critical temperature $T_c^\text{sf}$ for the superfluid transition. The 2D superfluid transition is in principle also determined by the BKT mechanism, where the topological defects are now vortices. For weak pairing, however, the mean-field BCS theory in fact 
 gives a good estimate of the transition temperature. We thus determine the critical temperature by solving the linearised gap equation 
 \begin{align}
\Delta_\bk = -\int \frac{d\bk'}{(2\pi)^2}\calV(\bk,-\bk')\Delta_{\bk'}\left[\frac{\tanh(\xi_{\bk'}/2T)}{2\xi_{\bk'} }- \frac{\calP}{2\xi_{\bk'}} \right ]. 
\label{gapeq}
\end{align}
Here $\Delta_\bk $ is the gap parameter and $\calV(\bk,-\bk')$ is the effective interaction between the quasiparticles in the stripe phase with energy dispersion $\xi_\bk$ measured from the Fermi surface. The details of this calculation are given in the methods section. 
The critical temperature obtained from this calculation is shown in Fig.~\ref{Tctheta} for $g = 1$ and for several tilting angles. This mean-field result 
gives an upper bound to the critical temperature, but since  $T_c^\text{sf}\ll \epsilon_F$ we expect that a more detailed BKT calculation yield only slightly smaller 
values. This should be contrasted with the melting of the stripes, where an estimate of the critical temperature from a vanishing stripe order would give a much higher value compared to the BKT calculation. This can be seen from Fig.\ \ref{BT}, which shows that the mean-field elastic coefficients remain large up to $T=0.12\epsilon_F$. Thus, it is crucial to use the BKT theory to analyse the stripe melting. 

Using a simple $p$-wave ansatz for the gap parameter $\Delta_\bk \simeq \Delta \cos\phi$, where $\phi$ is the polar angle of the wave vector $\bk$, one can 
obtain an approximate solution for the critical temperature as 
\begin{align}
T_c^\text{sf} \simeq C \epsilon_F e^{- {1}/ {\left [g\left (\frac 9 4 \sin^2\Theta-1\right )\right ]}},
\label{Ts}
\end{align}
where C is a constant related to an effective momentum cutoff in the integral in (\ref{gapeq}). We find that the data obtained from solving (\ref{gapeq}) numerically are in fact very well described by  (\ref{Ts}) with $C\simeq 0.4$.

\subsection{Quantum nematic phase for $\Theta=0$}
Figure  \ref{Tctheta} shows that the critical temperature for the stripe phase vanishes as $\Theta\rightarrow 0$. This is a direct consequence of the rotational symmetry rendering $B_\perp=0$ for $\Theta=0$. In this case, the system is no longer described by the XY model. Instead,  an appropriate expression for the elastic energy of stripe fluctuations  is~\cite{Toner1981}
\begin{align}
F_{\rm el}=\frac 1 2 B_\parallel\int d^2r[\lambda^2(\partial_x^2u)^2+(\partial_y u)^2],
\label{TonerNelsonE}
\end{align}
where $\lambda$ is a length comparable to the stripe spacing. Dislocations again play an important role in determining the finite temperature properties of the system described by (\ref{TonerNelsonE}). In contrast to the $\Theta>0$ case, however, single dislocations now have a  finite energy and can be thermally excited.  When the presence of the free dislocations is taken into account, a  system described by (\ref{TonerNelsonE}) is predicted to be in a nematic phase for $0<T<T_c^\text{n}$, and in an isotropic 
liquid phase for $T>T_c^n$~\cite{Toner1981}. In the nematic phase, the translational order exists only within a length scale $\xi_D$, which is determined by the density of the free dislocations. The stripe orientations, averaged over the length scale $\xi_D$, are however algebraically correlated. As a crude physical picture, one can think of the 
nematic phase as blobs of stripe order of area  $\sim\xi_D^2$, which are all oriented more or less in the same direction, but which are not positionally correlated with each other. The nematic phase is in this sense  analogous to the 2D hexatic phase of a crystal, which exhibits bond orientational order but no long-range translational order~\cite{Halperin1978,Nelson1979,NelsonBook}. A quantum hexatic phase was recently predicted to exist in 2D dipolar gases  for 
very strong coupling $g\gtrsim27 $~\cite{Bruun2014,Lechner2014}. The results presented here point out the intriguing possibility to realise a quantum version of the nematic phase with dipoles for smaller coupling strengths.
 We expect the critical temperature $T_c^\text{n}$ for the melting of the quantum nematic phase to scale as $B_\parallel$. However, a quantitative calculation of the critical temperature for the dipolar system requires knowledge of the parameter $\lambda$, whose determination is beyond our current theoretical framework. In Fig.~\ref{Tctheta}, we have indicated the critical temperature $T_c^n$ using a  somewhat smaller value than the bare $B_\parallel$ due to renormalisation effects.

\section{Discussion}
An important question concerns whether the critical temperature for the predicted quantum liquid crystal phases is within experimental reach.  
As an example, let us consider a recent experiment reporting the trapping of  chemically stable $^{23}$Na$^{40}$K molecules in their ground state close to quantum degeneracy.
The group obtained  an induced dipole moment of $d=0.8$Debye and a maximum 3D density of $n_\text{3D}=2.5\times 10^{11}$cm$^{-3}$~\cite{Park2015}. Estimating a corresponding 2D areal density as $n_0=n_\text{3D}^{2/3}$, these values correspond to $g\simeq0.57$. This coupling strength can be increased 
 by reaching a larger fraction of the permanent electric dipole moment of $^{23}$Na$^{40}$K, which is $2.72$Debye~\cite{Gerdes2011}, or by increasing the density of the gas. Since the critical temperature for the nematic and the stripe phases both scale as  $\sim (n_1/n_0)^2g\epsilon_F$, this indicates that the quantum liquid crystal physics discussed in this paper is within experimental reach, once dipolar gases can be cooled down significantly below their Fermi temperature. 

The formation of stripe  and superfluid order can be observed as correlation peaks in time-of-flight (TOF) experiments~\cite{Block2014,ZhigangWu2015}. 
One can also detect the stripes directly as density modulations, either after TOF or \emph{in-situ}, provided that the experimental resolution is sufficiently high. Observing the proliferation of dislocations would directly confirm the \emph{microscopic} mechanism behind the BKT transition. 

Finally, we would like to mention a recent  fixed note Monte-Carlo calculation which suggests that the striped phase is not the ground state  for $\Theta=0$
for any coupling strength~\cite{Matveeva2012}. We speculate that this result, 
which contradicts that of Refs.~\cite{Yamaguchi2010,Babadi2011,Sieberer2011,Block2012,Block2014,vanZyl,Parish2012}, is due to the
 approximate nature of the calculation combined with the fragility of the striped phase, which melts at any non-zero temperature for $\Theta=0$, as shown by our results. 

In summary, we analysed the phase diagram of a 2D dipolar gases, which exhibits stripe, nematic and supersolid phases corresponding to the breaking 
of translational, rotational and gauge symmetry.  For a non-zero tilting angle $\Theta$, the low energy degrees of freedom of the striped phase are described by an anisotropic 2D XY model. We calculated the stiffness constants corresponding to a rotation and a compression/expansion of the stripes microscopically. 
This should be contrasted with electron systems, where  such stiffness constants are often simply unknown parameters of the theory. 
The stripes were shown to melt via the Berezinskii-Kosterlitz-Thouless mechanism due to the proliferation of dislocations, and we obtained the melting temperature using the relevant renormalisation group equations. 
We also calculated the critical temperature of the supersolid phase. For $\Theta=0$, the striped phase is stable only at $T=0$, which melts into a nematic phase for arbitrarily small temperatures. Our analysis of the melting temperatures demonstrated that they should be within experimental reach. An observation of these phases would constitute a major breakthrough in our understanding of the interplay between liquid crystal and superfluid order in low-dimensional  many-body systems.

\section{Methods}
\subsection{Renormalisation group equations}
We calculate the  softening of the effective stiffness constant $B=\sqrt{B_\parallel B_\perp}$ due to the excitations of dislocation pairs using the well-known renormalisation group equations 
\begin{align}
\frac{d K^{-1}(l)}{dl} = 2\pi^3 y^2(l) \nonumber \\
\frac{d y(l)}{dl} = [2-\pi K(l)] y(l).
\label{RGeqn}
\end{align}
Here $K(l) = B(l)/T$ and  $y(l)=\exp[-E_c(l)/T]$ are the scale-dependent stiffness constant and dislocation fugacity respectively. They both decrease 
with increasing $l$ as the renormalisation due to dislocation pairs at larger length scales are included via the solution of (\ref{RGeqn}). 
The initial values of $K(0)$ and $y(0)$ are the bare (local) values unrenormalised by dislocation pairs, which we calculate microscopically as described in the text. At a critical temperature $T_c^\text{st}$, the long range renormalised elastic coefficient $B_R\equiv\lim_{l\rightarrow\infty}B(l)$ drops to zero by a sudden jump of $2T_c^\text{st}/\pi$. This disappearance of elastic rigidity signals the melting of the stripes. 

\subsection{Mean-field theory of stripe formation}
The mean-field Hamiltonian that takes into account the possibility of stripe formation with a wave vector $\bq$ is given by~\cite{Block2014} 
\beq
\hat \calH_{MF} =\sum_{\bk}\e_\bk\hat c_\bk^\dag \hat c_\bk +\sum_\bk[h_\bk\hat c^\dag_{\bk+\bq}\hat c_\bk + h.c.],
\label{Hdwp}
\eeq
where $\hat c_\bk^\dag$ creates a dipole with momentum $\bk$,  $\e_\bk$ is the single particle Hartree-Fock energy 
\beq
\e_\bk = \frac{k^2}{2m} + \frac{1}{A}\sum_{\bk'}[V(0)-V(\bk-\bk')]\la \hat c_{\bk'}^\dag \hat c_{\bk'}\ra,
\label{ek}
\eeq
and $h_\bk$ is a real off-diagonal element defined by
\beq
h_\bk = \frac{1}{A}\sum_{\bk'}[V(\bq)-V(\bk-\bk')]\la \hat c^\dag_{\bk'}\hat c_{\bk'+\bq}\ra.
\label{hk}
\eeq
The quasi-2D interaction in Fourier space is obtained by averaging  the interaction over the harmonic oscillator ground state in the $z$ direction. This gives (up to an irrelevant constant term)~\cite{Fischer2006}
\begin{align}
V(\bk) &\simeq -g\frac{3\pi^2 k}{2mk^0_F}\left ( \cos^2\Theta-\sin^2\Theta\cos^2\varphi \right ),
\label{V2Df1}
\end{align}
where $\varphi$ is the polar angle of $\bk$. We diagonalise the mean-field Hamiltonian by generalising the method described in Refs.~\cite{Block2014,ZhigangWu2015} to an  arbitrary  stripe vector $\bq$. This yields the Hamiltonian 
\begin{align}
\hat \calH_{MF} = \sum_{j\bar\bk} \varepsilon_{j\bar\bk} \hat \gamma^\dag_{j\bar\bk} \hat \gamma_{j\bar\bk}.
\end{align}
Here $\hat \gamma_{j\bar\bk} = \sum_\bG U_{j,\bar\bk+\bG}\hat c_{\bar\bk + \bG}$ annihilates a quasiparticle with energy $\varepsilon_{j\bar\bk}$, where 
$j=1,2,\cdots$ is the band index, $\bG = l \bq, l =0,\pm 1,\cdots$ is the reciprocal lattice vector and $\bar\bk$ is restricted to the first Brillouin zone of the 1D periodic potential set up by the stripes. We can then calculate the mean-field free energy as $F_\text{MF}(\bq) = \Omega_\text{MF} + \mu N$, where $\Omega_\text{MF}$ is the mean-field thermodynamic potential given by (\ref{OmegaMF}) and $N=\sum_{j,\bar \bk} f_{j\bar \bk} $ with $f_{j\bar \bk}=[\exp\beta(\varepsilon_{j\bar\bk}-\mu)+1]^{-1}$. The interaction energy is most easily calculated using 
\begin{align}
E_{\text{MF}} = \sum_{j,\bar \bk} \varepsilon_{j\bar\bk}f_{j\bar \bk} =  E_{\text{kin}} + 2  E_{\text{int}}
\end{align} 
where $E_{\text{kin}} = \sum_\bk \la \hat c^\dag_\bk \hat c_\bk\ra k^2/2m$ the kinetic energy. 

\subsection{BCS theory of the superfluid transition}
To explore  superfluid pairing within the stripe phase, we use BCS theory with the quasiparticle Hamiltonian $\hat\calH_{\rm BCS}=\hat \calH_{\rm MF}+\hat \calH_{\rm P}$. Here,   
 \begin{align}
\hat \calH_{\rm P} = \sum_{{jj' 
 \bar\bk\bar\bk'}}\frac{\calV_{j'j}(\bar\bk',-\bar\bk)}2\la \hat \gamma^\dag_{j'\bar\bk'}\hat \gamma_{j',-\bar\bk'}^\dag\ra\hat \gamma_{j\bar\bk}\hat \gamma_{j,-\bar\bk} + \rm{h.c.} \nn
\end{align}
describes pairing between the time-reversed quasiparticles, interacting via 
\begin{gather}
\calV_{j'j}(\bar\bk',-\bar\bk) = \sum_{\bG\bG'\tilde\bG \tilde \bG'}\delta_{\bG-\bG',\tilde \bG'-\tilde\bG}U^*_{j',\bar\bk'+\bG'} U^*_{j',-\bar\bk'+\tilde\bG'}\nonumber\\
\times U_{j,-\bar\bk+\tilde\bG}U_{j,\bar\bk+\bG}V\left(\bar\bk-\bar\bk'+\bG-\bG'\right ).
\label{Vmatrdw2}
\end{gather}
 To derive a gap equation that is amenable to a partial wave expansion, we switch to the ``extended zone scheme", whereby a single 
 particle state $\psi_{j\bar\bk}(\brho)$ in the $j$'th band in the first BZ is mapped onto a state $\psi_{\bk}(\brho)$ in the $j$'th BZ in the standard way~\cite{ZhigangWu2015}, where the vector $\bk$ is now unrestricted. The effective pairing interaction $\calV_{j'j}(\bar\bk',-\bar\bk) $ shall be denoted by  $\calV(\bk,-\bk')$ and quasi-particle dispersion $\varepsilon_{j\bar\bk}$ by $\varepsilon_\bk$. Pairing between time-reversed quasiparticles gives rise to the gap parameter 
$\Delta_{\bk}  \equiv \sum_{\bk'} \calV(\bk,-\bk')\la\hat \gamma_{-\bk'}\hat \gamma_{\bk'} \ra$, which satisfies the finite temperature gap equation  
\beq
\Delta_\bk = -\int\frac{d\bk'}{(2\pi)^2}\calV(\bk,-\bk')\Delta_{\bk'}\left [\frac{\tanh(E_{\bk'}/2T)}{2E_{\bk'}} -\frac{\calP}{2\xi_{\bk'}}\right ].
\label{gapeq2}
\eeq
Here $\xi_\bk = \varepsilon_\bk-\mu$ and $E_\bk=\sqrt{\xi_{\bk}^2+|\Delta_\bk|^2}$, where the chemical potential $\mu$ is approximated by the value in the stripe phase. The Cauchy principal value term $\calP/2\xi_{\bk'}$ in (\ref{gapeq2}) renders the gap equation well defined with no need for a high momentum cut-off. At temperatures in the vicinity of the superfluid transition, the linearisation of the above gap equation yields (\ref{gapeq}) we use in the main text. Equation (\ref{gapeq}) can be solved by the method of partial wave expansion described in Ref.~\cite{ZhigangWu2015}. Finally we determine the transition temperature by gradually increasing $T$ in the gap equation until it ceases to admit finite solutions.

\section{Acknowledgement}
G.M.B. would like to acknowledge the support of the Hartmann Foundation via grant A21352 and the Villum Foundation via grant VKR023163.

\section{Additional information}
{\bf Competing financial interests}: The authors declare no competing financial interests.

\end{document}